\documentclass[%
 reprint,
 amsmath,amssymb,
 aps,
]{revtex4-1}

\usepackage{graphicx}% Include figure files
\usepackage{dcolumn}% Align table columns on decimal point
\usepackage{bm}% bold math
\begin{document}
\preprint{APS/123-QED}

\title{Analytical soliton solution for the Landau-Lifshitz equation of \\one dimensional magnonic crystal}% Force line breaks with \\
\author{D. Giridharan}

\author{M. Daniel*}
\affiliation{Centre for Nonlinear Dynamics, School of Physics, Bharathidasan University, Tiruchirapalli 620 024, India
 }%
\author{P. Sabareesan}
\affiliation{%
Centre for Nonlinear Science and Engineering, School of Electrical and Electronics Engineering, Sastra University, Thanjavur - 613401, India
}%

\date{\today}% It is always \today, today,
             %  but any date may be explicitly specified

\begin{abstract}
\hspace{1cm}Nonlinear localized magnetic excitations in one dimensional magnonic crystal is  investigated under periodic magntic field. The governing Landau-Lifshitz equation  is transformed into variable coefficient nonlinear Schrodinger equation(VCNLS) using sterographic projection. The VCNLS equation is in general nonintegrable, by using painleve analysis necessary conditions for the VCNLS equation to pass Weiss-Tabor-Carnevale (WTC) Painleve test are obtained. A sufficient integrability condition is obtained by further exploring a transformation, which can map the VCNLS equation into the well-known standard nonlinear Schrodinger equation. The transformation built a systematic connection between the solution of the standard nonlinear Schrodinger equation and VCNLS equation. The results shows the excitation of magnetization in the form of soliton has spatial period exists on the background of spin Bloch waves. Such solution exisits only certain constrain conditions on the coefficient of the VCNLS equation are satisfied. The analytical results suggest a way to control the dynamics of magnetization in the form of solitons by an appropriate spatial modulation of the nonlinearity coefficient in the governing VCNLS equation which is determined by the ferromagnetic materials which forms the magnonic crystal.
%By using, painleve analysis integrability conditions are obtained for the VCNLS equation. Futhermore we use a transformation, which can map the VCNLS equation into the well-known standard nonlinear schrodinger equation. The transformation bulits a systematic connection between the solution of the standard nonlinear Schrodinger equation and VCNLS equation. The finding of the transformation has significant contribution to understand the nonlinear magnetization dynamics in magnonic crysatal and also study the impact of material parameter variation on the localized excitation of magnetism in one dimensional magnonic crystal. The results shows the existence of spatially localized modulated amplitude ferromagnetic gap solitons which describes the nonlinear magnetic excitations in one dimensional magnonic crystal. The analytical results suggest a way to control the dynamics of magnetization in the form of solitons by an appropriate spatial modulation of the nonlinearity strength in the governing VCNLS equation. Also It is found that the amplitude of the soliton solution is depend on the material parameter and by changing the ferromagnetic materials which forms the magnonic crystal, it is possible to tune the spatially modulated amplitude of the ferromagnetic gap soliton.  
\begin{description}
\item[Usage]
Secondary publications and information retrieval purposes.
\item[PACS numbers]
May be entered using the \verb+\pacs{#1}+ command.
\item[Structure]
You may use the \texttt{description} environment to structure your abstract;
use the optional argument of the \verb+\item+ command to give the category of each item. 
\end{description}
\end{abstract}
\pacs{Valid PACS appear here}% PACS, the Physics and Astronomy
                             % Classification Scheme.
%\keywords{Suggested keywords}%Use showkeys class option if keyword
                              %display desired
\maketitle
%\tableofcontents
%\setstretch{1.5}
%\vskip 10pt
%\textbf{\large {Introduction:}}\\
\section*{I. Introduction}
\hspace{1cm}The study of nonlinear magnetic excitations in terms of solitary waves and solitons in ferromagnetic systems have attracted much interest in the past several years[1-7]. The results reveal that the dynamics is governed by Landau-Lifshitz equation which can be mapped to Nonlinear Schrodinger(NLS) family of equations[8]. In recent years, the studies on nonlinear systems with spatial periodicity has become a great topic of interest[9]. BEC in optical lattices[10,11], solitons in Photonic lattices [12] and periodic magnetic systems[13-15] etc., are the typical models among them. Motivated by these considerations, in the present paper we investigate the nature of excitation of magneization in one dimensional magnonic crysatal. Magnonic crysatal is a medium with spatially periodic variation of their magnetic properties in a definite direction. In the linear regime, observation of frequency band gap is well studied problem. The fundamental feature of periodic magnetic structures is energy band gap in their spectrum of spin waves. The band gap represents a range of energy values in which spin-wave excitations are forbidden from propagating. The theoritical and experimental studies related to magnonic crystal mostly is devoted to linear phenomena. The investigation of propagation of soliton in magnonic crystal are insufficent, only few specific studies in the field that shows the experimental and numerical simulation results based on one dimensional NLS equation. The bright and dark solitons were observed in yittrium iron garnet films with artificial periodicity[13-15]. Morozove et.al[16], investigated the features of formation of the soliton that are similar to bragg solitons in the ferromagnetic one dimensional periodic structure using coupled mode theory. He et.al[17], studied the modulation instability and gap solitons in ferromagnetic films under periodic magnetic field using multiscale expansion method. The earlier studies are based on a homogeneous ferromagnetic films and achieve periodicity by varying the thickness of the films or by applying spatially periodic magnetic field[16,17]. Here in this present study, we consider an infinite one dimensional magnonic crystal formed by periodic array of distinct elements and study the impact of material parameter variation on the localized excitation of magnetization under periodic magnetic field.. The paper is organized as follows. In Sec.II we consider the one dimensional magnonic crystal model under periodic magnetic field and derive the dynamical equation. In Sec.III The governing VCNLS equation is analyzed through painleve analysis to obtain integrability conditions and it is mapped into standard NLS equation using suitable transformation. In Sec.IV The results are presented.
\section*{II. Model and Dynamical equation}
\hspace{1cm}We consider an infinite one dimensional magnonic crystal represented by a system of alternating uniform ferromagnetic layers of two different materials A and B as shown in Fig.(1). The layers have different values of exchange length parameter,$J_{ex}$ and saturation magnetization,$M_{S}$ throughtout the sample. Let $J_{ex,A}$ and $J_{ex,B}$ be the exchange length for the ferromagnetic materials A and B respectively. Let $M_{S,A}$ and $M_{S,B}$ be the saturation magnetization for the ferromagnetic materials A and B respectively. Here z-axis is chosen normal to the plane of the layers. The equation of motion of the magnetization in the 1D-magnonic crystal is governed by the following Landau-Lifshitz(LL) equation[18].
\begin{equation}
\frac{ \partial \vec M(\vec r,t)}{\partial t} = -\gamma \vec M(\vec r,t) \times \vec H_{eff}(\vec r,t),
\end{equation}
where $\gamma$ is the gyromagnetic ratio and  $\vec H_{eff}$ denotes the effective field. In general, the effective field $\vec H_{eff}$ is the sum of several components includes the applied  field, anisotropy field, demagnetization field and exchange field which are all dependent on space.
\begin{equation}
\vec H_{eff}=\vec H_{0}+\vec H_{ani}+\vec H_{d}+\vec H_{ex}.
\end{equation}
The first component $\vec H_{0}$  is the applied magnetic field which is inhomogeneous in space and applied along z-direction. The next component is the anistopy field, $\vec H_{ani}$ which is given by
\begin{equation} 
\vec H_{ani}=\beta(x)M_{z}\hat{z}. 
\end{equation}
The another component  $\vec H_{d}$ arises entirely from the demagnetizing field that corresponds to shape anisotropy. 
\begin{equation}
\vec H_{d}(\vec r, t)=\lambda(N_{x}M_{x}\hat{x} + N_{y}M_{y}\hat{y} +N_{z}M_{z}\hat{z} ), 
\end{equation}
In the case of an xy-film, the demagnetization field is given by $\vec H_{d}=\lambda M_{z}\hat{z}$.\\
Where $\lambda = -1$.\\
The exchange field $\vec H_{ex}$ given by[19],
\begin{equation}
\vec H_{ex}=J_{ex}(x) \vec \nabla^{2} \vec M(\vec r,t)),
\end{equation}
where $J_{ex}(x)={\frac{2A(x)}{\mu_{o}M_{S}(x)^{2}}}$ is the exchange length, $A(x)$ is the exchange constant and $M_{S}(x)$ is the saturation magnetization. Thus, the total effective field $\vec H_{eff}$ takes the form,
\begin{equation}
\vec H_{eff}=H_{0}(x) \hat{z}+\beta(x)M_{z}\hat{z}+\lambda M_{z}\hat{z}+ J_{ex}(x) \vec \nabla^{2} \vec M(\vec r,t)).
\end{equation}
\begin{center}
\begin{figure}
\includegraphics[width=0.65\columnwidth]{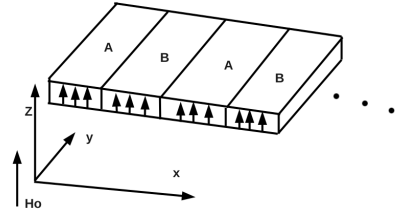}
\caption{Schematic drawing of one dimensional magnonic crystal and its coordinate system. A and B are two ferromagnetic materials.}
\end{figure}
\end{center}
Upon using the above expression for the effective field in Eq.(1), we get
%\begin{equation}
%\frac{ \partial \vec M(\vec r,t)}{\partial t} = -\gamma \vec M \times [ H_{0}(x) \hat{z}+\beta(x)M_{z}\hat{z}+\lambda M_{z}\hat{z}+ J_{ex}(x) \vec \nabla^{2} \vec M(\vec r,t)],  M_{x}^2 + M_{y}^2 + M_{z}^2 = M_{s}(x)^2.
%\end{equation}
\begin{eqnarray}
\frac{ \partial \vec M(\vec r,t)}{\partial t} &=& -\gamma \vec M \times [ H_{0}(x) \hat{z}+\beta(x)M_{z}\hat{z}+\lambda M_{z}\hat{z}\quad\nonumber
\\
&&+ J_{ex}(x) \vec \nabla^{2} \vec M(\vec r,t)],\\\nonumber
&& M_{x}^2 + M_{y}^2 + M_{z}^2 = M_{s}(x)^2.
\end{eqnarray}
$J_{ex}(x)$, $M_{S}(x)$ and $\beta(x)$ are periodic functions with period equal to the magnonic crystal lattice constant a.\\
The exchange length, $J_{ex}(x)$ is represented as,
\[ J_{ex}(x+a) = J_{ex}(x) = %
\begin{cases}
J_{ex,A} &   0  \leq x < a/2, \\
J_{ex,B}              & a/2  \leq x < a.
\end{cases}
\]
The saturation magnetization, $M_{s}(x)$ is represented as,
 \[ M_{S}(x+a) = M_{S}(x) = %
\begin{cases}
M_{s,A} &  0  \leq x < a/2, \\
M_{s,B}              &  a/2 \leq x < a.
\end{cases}
\]
Similarly, the anisotropy constant, $\beta(x)$ is represented as,
 \[ \beta(x+a) = \beta(x) = %
\begin{cases}
\beta_{A} &  0  \leq x < a/2, \\
\beta_{B}              &  a/2 \leq x < a.
\end{cases}
\]
Since the Landau-Lifshitz equation is a continous equation, the material parameters should be represented in continous form. To make its continous which are expanded in a Fourier series in the form,\\
\begin{equation}
J_{ex}(x)=c_{o}+\sum_{n=1} c_{n} cos(\frac{2n \pi x}{a}) +\sum_{n=1}  d_{n}sin(\frac{2n \pi x}{a}). 
\end{equation}\\
After evaluating the coefficients of the Fourier series, we get the function $J_{ex}(x)$ in the form,\\
\begin{eqnarray}
J_{ex}(x)&=&(\frac{J_{ex,A}+J_{ex,B}}{2})+\sum_{n=1}(\frac{\Delta J}{n \pi})(1-(-1)^{n}))\quad\nonumber
\\
&&\times sin(\frac{2n \pi x}{a}),
\end{eqnarray}
\begin{equation}
J_{ex}(x)=J_{av}+\sum_{n=1}(\frac{\Delta J}{n \pi})(1-(-1)^{n}))sin(\frac{2n \pi x}{a}),
\end{equation}
The saturation magnetization $M_{s}(x)$ is represented as,
\begin{eqnarray}
M_{S}(x)&=&(\frac{M_{S,A}+M_{S,B}}{2})+\sum_{n=1}(\frac{\Delta  M_{S}}{n \pi})(1-(-1)^{n}))\quad\nonumber
\\
&&\times sin(\frac{2n \pi x}{a}),
\end{eqnarray}
\begin{equation}
M_{S}(x)=M_{S,av}+ \sum_{n=1}(\frac{\Delta  M_{S}}{n \pi})(1-(-1)^{n}))sin(\frac{2n \pi x}{a}).
\end{equation} 
Similarly, the anisotropy constant $\beta(x)$ is represented as,
\begin{equation}
\beta(x)=(\frac{\beta_{A}+\beta_{B}}{2}) + \sum_{n=1} (\frac{\Delta \beta}{n \pi})(1-(-1)^{n}))sin(\frac{2n \pi x}{a}),
\end{equation}
\begin{equation}
\beta(x)= \beta_{av}+ \sum_{n=1} (\frac{\Delta \beta}{n \pi})(1-(-1)^{n}))sin(\frac{2n \pi x}{a}).
\end{equation}
where, $\Delta J= J_{ex,A}-J_{ex,B}$, $\Delta M= M_{S,A}-M_{S,B}$, $\Delta \beta= \beta_{A}-\beta_{B}$ and $J_{av}$, $M_{S,av}$ and $\beta_{av}$ represents the average exchange length, saturation magnetization and anisotropy constant value of two ferromagnetic materials A and B respectively. The material parameters varies gradually at the interface between the two ferromagnetic materials and the average values of the material parameters represents, a value of material parameter of the periodic ferromagnetic system at the exact centre interface point between the two ferromagnetic materials.\\ 
The LL equation is a vector nonlinear partial differential equation, it is difficult to solve in its original form. By using sterographic projection, we transform the LL equation into a nonlinear equation of a complex function. In the component form Eq.(7) becomes,
          %\begin{equation}
          %\frac{ \partial M_{x}}{\partial t} = -\gamma \{H_{0}(x)M_{y} +\beta(x)M_{z}M_{y}+ \lambda M_{z} M_{y} + J_{ex}(x) ( M_{y}\nabla^{2} M_{z} - M_{z}\nabla^{2} M_{y})\},
          %\end{equation}
          %\begin{equation}
          %\frac{ \partial M_{y}}{\partial t} = -\gamma \{-H_{0}(x)M_{x}-\beta(x)M_{z}M_{x}-\lambda M_{z} M_{x} - J_{ex}(x) ( M_{x}\nabla^{2} M_{z} - M_{z}\nabla^{2} M_{x})\},
          %\end{equation}
          %\begin{equation}
          %\frac{ \partial M_{z}}{\partial t} = -\gamma \{{J_{ex}(x) ( M_{x}\nabla^{2} M_{y} - M_{y}\nabla^{2} M_{x})}\}.
          %\end{equation}
\begin{eqnarray}
\frac{ \partial M_{x}}{\partial t} &=& -\gamma \{H_{0}(x)M_{y} +\beta(x)M_{z}M_{y}+ \lambda M_{z} M_{y}\quad\nonumber \\
&&+ J_{ex}(x) ( M_{y}\nabla^{2} M_{z} - M_{z}\nabla^{2} M_{y})\},
\end{eqnarray}
\begin{eqnarray}
\frac{ \partial M_{y}}{\partial t} &=& -\gamma \{-H_{0}(x)M_{x}-\beta(x)M_{z}M_{x}-\lambda M_{z} M_{x}\quad\nonumber \\
&&- J_{ex}(x) ( M_{x}\nabla^{2} M_{z} - M_{z}\nabla^{2} M_{x})\},
\end{eqnarray}
\begin{eqnarray}
\frac{ \partial M_{z}}{\partial t} &=& -\gamma \{{J_{ex}(x) ( M_{x}\nabla^{2} M_{y} - M_{y}\nabla^{2} M_{x})}\}.
\end{eqnarray}
Defining,
	\begin{equation}
        \psi(x,t) = \frac{M_{x} + i M_{y}}{M_{S}(x)},
	\end{equation}
        \begin{equation}
        \psi^*(x,t) = \frac{M_{x} - i M_{y}}{M_{S}(x)},
        \end{equation}
where $\psi$ is a complex variable and then we have,
\begin{equation}
        m_{z}(x,t) = (1-\lvert\psi  \rvert^2)^{1/2}. 
\end{equation}
Considering small deviations of magnetization from the equilibrium direction corresponding to $\lvert\psi  \rvert^2 <<1 $  and under the long wavelength approximation by keeping only the nonlinear terms of magnitude $\lvert\psi  \rvert ^2 \psi$[20] we obtain, 
%\begin{small}
%\begin{equation}
 %i \frac{ \partial \psi}{\partial t} =\left(\frac{J_{ex}(x)}{J_{av}}\right)\left(\frac{M_{S}(x)}{M_{S,av}}\right)\frac{\partial^{2} \psi}{\partial x^2}-\frac{1}{2}\left(1-\beta(x)\right)\left(\frac{M_{S}(x)}{M_{S,av}}\right) \lvert \psi \rvert^2 \psi-\left(\left(\frac{H_{0}(x)}{M_{S,av}}\right)-\left(1-\beta(x)\right)\left(\frac{M_{S}(x)}{M_{S,av}} \right)\right)\psi
%\end{equation}
%\end{small}
\begin{small}
\begin{eqnarray}
 i \frac{\partial \psi}{\partial t} &=& \left(\frac{J_{ex}(x)}{J_{av}}\frac{M_{S}(x)}{M_{S,av}}\right)\frac{\partial^{2} \psi}{\partial x^2}-\frac{1}{2}\left(1-\beta(x)\right)\left(\frac{M_{S}(x)}{M_{S,av}}\right)\lvert \psi \rvert^2 \psi\nonumber \\
&&-\left(\left(\frac{H_{0}(x)}{M_{S,av}}\right)-\left(1-\beta(x)\right)\left(\frac{M_{S}(x)}{M_{S,av}}\right)\right)\psi
\end{eqnarray}
\end{small}
%Here the temporal and spatial coordinates are rescaled by 1/$(\gamma  M_{S,av})$ and $\sqrt(J_{ex,av})$ respectively.\\ \\
%Let $f(x)=\left(\frac{J_{ex}(x)}{J_{av}}\right) \left(\frac{M_{S}(x)}{M_{S,av}}\right) $, $g(x)=\left(1-\beta(x)\right)\left(\frac{M_{S}(x)}{M_{S,av}}\right)$, $h(x)=\left(\frac{H_{0}(x)}{M_{S,av}}\right)-\left(1-\beta(x)\right) \left(\frac{M_{S}(x)}{M_{S,av}} \right)$
Here the temporal, and spatial coordinates are rescaled by $t_{o}=1/(\gamma  M_{S,av})$ and $l_{o}=\sqrt(J_{av})$ respectively.\\ 
Let 
\begin{center}
{\small$f(x)=\left(\frac{J_{ex}(x)}{J_{av}}\right) \left(\frac{M_{S}(x)}{M_{S,av}}\right) $}, {\small$g(x)=\left(1-\beta(x)\right)\left(\frac{M_{S}(x)}{M_{S,av}}\right)$}
\end{center}
\begin{center} 
{\small$h(x)=\left(\frac{H_{0}(x)}{M_{S,av}}\right)-\left(1-\beta(x)\right)\left(\frac{M_{S}(x)}{M_{S,av}}\right)$}
\end{center}
Then the Eq.(21) becomes,
\begin{equation}
i \frac{ \partial \psi}{\partial t} - f(x)\frac{\partial^{2} \psi}{\partial x^2} + \frac{1}{2} g(x)\lvert \psi \rvert^2 \psi + h(x)\psi =0
\end{equation}
The Eq.(22) is the Nonlinear Scr\"{o}dinger equation with  variable coefficients. \\When $f(x)=g(x)=h(x)=$ constant Eq.(22) reduces to completely integrable nonlinear schr\"{o}dinger equation which admits N-soliton solutions[21].\\\\
In the absence of cubic term, the above Eq.(22) is a linear periodic system and admits Bloch wave solutions. As mentioned earlier, in the linear studies observation frequency band gap is well studied problem. The bandgap represents the range of energy values in which spin wave excitations are forbidden from propagating. From the fabrication point of view, experimental studies were performed on periodic structures composed of only one constituent magnetic material and achieve one dimensional periodicity in homogeneous ferromagnetic films which includes micron size shallow grooves etched on yttrium iron garnet(YIG) films[22], one dimensional array of micron size metal stripes on YIG films[23] and by applying a periodic magnetic field of spatially varying strength[24]. Wang et.al.,[25,26] have achievd first bicomponent magnonic crystal experimentally which exhibit well defined frequency bandgaps. The experimental results revealed that the tunability of the magnonic band gaps can be achieved by varying the width of the component stripes or by varying the materials.\\
In the presence of cubic term, It is completely nonlinear problem. The above VCNLS equation is in general nonintegrable, to solve the above equation analytically, in the next section integrability conditions are obtained by using painleve analysis. The integrability conditions are expressed in terms cofefficients of the VCNLS equation. The above equation admits soliton solutions only when the integrability conditions are satisfied.  
%%%%%%%%%%%%%%%%%%%%%%%%%%%%%%%%%%%%%%%%%%%%%%%%%%
\section*{III. Painleve analysis and integrability conditions}
\hspace{1cm}Several tools such as Painleve analysis[27], Lax pair[28] and similarity transformation techniques[29] are available to solve VCNLS equation to obtain analytical solutions. Our analysis is based on the Painleve test for partial differential equations i.e., the Weiss-Tabor-Carnevale (WTC) test, which has been found to be a successful tool for investigate the integrability of partial differential equations. In this section, we use WTC test to obtain an integrability condition for the VCNLS equation and then under this condition, we look for a transformation which converts the Eq. (22) to the standard NLS equation. In order to perform conveniently, we rewrite the Eq. (22) and its complex conjugate by replacing $\psi$ by a and $\psi^*$ by b obtain,
\begin{subequations}
\begin{align}
i \frac{ \partial a}{\partial t} - f(x)\frac{\partial^{2} a}{\partial x^2} + \frac{1}{2} g(x) a^2 b + h(x)a =0
\end{align} 
\begin{align}
-i \frac{ \partial b}{\partial t} - f(x)\frac{\partial^{2} b}{\partial x^2} + \frac{1}{2} g(x) b^2 a + h(x)b =0
\end{align} 
\end{subequations}
Where $f(x)$, $g(x)$ and $h(x)$ are real functions.\\ \\
The next step is to seek solution in the form of Laurent series,\\
\begin{subequations}
\begin{align}
a(x,t)=\large \Sigma_{j=0}^{\infty } a_{j}(x,t)\phi^{\alpha + j}(x,t) 
\end{align}
\begin{align}
b(x,t)=\large \Sigma_{j=0}^{\infty } b_{j}(x,t)\phi^{\alpha + j}(x,t) 
\end{align}
\end{subequations} 
Where $a_0$, $b_0$ $\neq$ 0 and $a_j$, $b_j$ and $\phi(x,t)$ are analytic functions. $\alpha$ and $\beta$ are negative integers to be determined from the leading order analysis.\\
%%%%%%%%%%%%%%%%%%%%%%%%%%%%%%%%%%%%%%%%%%%%%%%%%%%%%%%%%%%%%%%%%%%%%%%%%%%%%%%%%%%%%%%%%%%%%%%%%%%%%%%%%%%%%
Atr j=4, we obtain integrability conditions,
\begin{equation}
\left(\frac{f_x}{f} + 2\frac{g_x}{g}\right) =0
\end{equation}
i.e., 
\begin{equation}
f(x) = \frac{k}{g(x)^2} 
\end{equation}
\begin{equation}
h(x) = \left(\frac{k}{2}\right) \left[\frac{g_{xx}}{g^3} - \frac{3}{2} \frac{g_{x}^{2}}{g^4} \right] 
\end{equation}
Where k is an integration constant. By employing, Painleve method for the governing VCNLS Eq.(22) and we obtain integrability conditions.\\ 
Here,
\begin{center}
{\small$f(x)=\left(\frac{J_{ex}(x)}{J_{av}}\right) \left(\frac{M_{S}(x)}{M_{S,av}}\right) $}, {\small$g(x)=\left(1-\beta(x)\right)\left(\frac{M_{S}(x)}{M_{S,av}}\right)$}
\end{center}
\begin{center} 
{\small$h(x)=\left(\frac{H_{0}(x)}{M_{S,av}}\right)-\left(1-\beta(x)\right)\left(\frac{M_{S}(x)}{M_{S,av}}\right)$}
\end{center}
%Here,$f(x)=\left(\frac{J_{ex}(x)}{J_{av}}\right) \left(\frac{M_{S}(x)}{M_{S,av}}\right) $, $g(x)=\left(1-\beta(x)\right)\left(\frac{M_{S}(x)}{M_{S,av}}\right)$, $h(x)=\left(\frac{H_{0}(x)}{M_{S,av}}\right)-\left(1-\beta(x)\right) \left(\frac{M_{S}(x)}{M_{S,av}} \right)$\\ \\
The difference in anisotropy constant values $\beta$  between the constituent materials tends to slight and the effect of this inhomogenity is minor. Hence assume this component to be negligible.\\ 
Then,
\begin{center}
{\small$f(x)=\left(\frac{J_{ex}(x)}{J_{av}}\right) \left(\frac{M_{S}(x)}{M_{S,av}}\right) $}, {\small$g(x)=\left(\frac{M_{S}(x)}{M_{S,av}}\right)$}
\end{center}
\begin{center} 
{\small$h(x)=\left(\frac{H_{0}(x)}{M_{S,av}}\right)-\left(\frac{M_{S}(x)}{M_{S,av}}\right)$}
\end{center}
%Then, $f(x)=\left(\frac{J_{ex}(x)}{J_{av}}\right) \left(\frac{M_{S}(x)}{M_{S,av}}\right) $,~~~~ $g(x)=\left(\frac{M_{S}(x)}{M_{S,av}}\right)$,~~~ $h(x)=\left(\frac{H_{0}(x)}{M_{S,av}}\right) - \left(\frac{M_{S}(x)}{M_{S,av}}\right)$.\\ \\
Substuting this in eq.(37) and eq.(38) we get the integrability conditions as,\\
\begin{equation}
\left(\frac{J_{ex}(x)}{J_{av}}\right) = \left(k\frac{M_{S,av}^3}{M_{S}^3(x)}\right)
\end{equation}
and
\begin{equation}
\left(\frac{H_{0}(x)}{M_{S,av}}\right) - \left(\frac{M_{S}(x)}{M_{S,av}}\right)=M_{S,av}^2\left(\frac{-3M_{S}^{'2}(x) + 2M_{S}(x)M_{S}^{''}(x) }{M_{S}^{4}(x)}\right)
\end{equation}
i.e.,\\
\begin{equation}
\frac{H_{0}(x)}{M_{S,av}}= \left(\frac{M_{S}(x)}{M_{S,av}}\right)+M_{S,av}^2 \left(\frac{-3M_{S}^{'2}(x) + 2M_{S}(x)M_{S}^{''}(x) }{M_{S}^{4}(x)}\right)
\end{equation}\\
The above integrability conditions are consistent with Ref.[10] and Ref.[27]. The exchange constant, $J_{ex}(x)$ which is inhomogeneous in space related to the saturation magnetization $M_{S}(x)$ is given by Eq.(39) and  it is first integrability conditions for  Eq. (22) to be integrable. From Eq. (41) it is noted that form of the periodic applied magnetic field is determined by the ferromagnetic materials which forms the magnonic crystal. The applied magnetic field is periodic in space and its periodicity forms a periodic potential for the spin waves. The form of the potential is related to saturation magnetization $M_{S}(x)$ which is given by Eq.(41) and it is second integrability conditions for  Eq. (22) to be integrable.\\ 
Further, we have to construct the solution of VCNLS equation by using a transformation which converts the Eq. (22) into a standard NLS equation. We look for the transformation of the form[10],
\begin{equation}
\psi(x,t) = r(x) q(X(x),T(t))
\end{equation}
where $X=X(x)$ and $T=T(t)$ and $r(x)$ are the real functions to be determined.\\
Substuting Eq. (42) into Eq. (22), We get set of following equations
\begin{equation}
f(x)r_{xx} + h(x)r(x) = 0.
\end{equation}
\begin{equation}
2r_{x}X_{x} + r(x)X_{xx} = 0.
\end{equation}
\begin{equation}
T_{t} = f(x)X_{x}^{2} = g(x)r(x)^{2}
\end{equation}
and also,
\begin{equation}
iq_{T} - q_{XX} + \frac{1}{2}\lvert q \rvert^2 q  = 0.
\end{equation}
By using the integrability conditions, the above equations are solved and gives
\begin{align}
r(x) = \sqrt{\frac{1}{g(x)}}
\end{align}
\begin{align}
X(x) = \int{g(x)} dx ~~~~~~~~~~~~and~~~~~~~~~~~ T(t) = t
\end{align}
Then, we obtain solution of Eq. (22) by known solution of standard NLS equation, $q(X,T)$.
\begin{equation}
\psi(x,t) = \sqrt{\frac{1}{g(x)}} q(X(x),T(t))
\end{equation}
Here, q(X,T) is the solution of standard NLS equation Eq.(46) and there exists several methods to solve the standard NLS equation such as classical IST, DBT, Hirota bilinear method etc.,. In this paper, Hirota bilinear method is used to construct the dark one soliton solution. 
\begin{eqnarray}
q(X(x),T(t))&=&\frac{1}{\sqrt{2}}[C1 - 2iC2\tanh(C2(X-C1T))]\nonumber\\
&&\times \exp{[\frac{-i}{2}(C1^2 + 4C2^2)T]}  
\end{eqnarray}
\section*{V. Results}
From Eq.(49), we obtain
\begin{small}
\begin{eqnarray}
\psi(x,t) &=& \frac{1}{\sqrt{2g(x)}}[C1 - 2iC2\tanh(C2(X-C1T))]\nonumber\\
&&\times \exp{[\frac{-i}{2}(C1^2 + 4C2^2)T]}  
\end{eqnarray}
\end{small}
where the parameters C1 and C2 corresponds to the velocity and depth of the dark soliton.	
\begin{equation}
g(x)= 1 + \sum_{n=1} \left(\frac{M_{s,A}-M_{s,B}}{M_{S,av} n \pi}\right)(1-(-1)^{n}))sin(\frac{2n \pi x}{a}).
\end{equation}
where $M_{S,A}$ and $M_{S,B}$ are the saturation magnetization for the ferromagnetic materials A and B respectively. $M_{S,av}= \left(\frac{M_{S,A}+M_{S,B}}{2}\right)$ is the average saturation magnetization value of two ferromagnetic materials A and B. The material parameter $M_{s}(x)$ varies smoothly at the interface between the two ferromagnetic materials and the average value represents the saturation magnetization value of the periodic ferromagnetic system at the exact centre interface point between the two ferromagnetic materials.\\ 
As we mentioned earlier, Landau-Lifishitz equation is a continous equation which describes the equation of motion of the magnetization in a ferromagnetic medium. Here we consider a periodic ferromagnetic system in which the material parameter varies periodically. In order to make it continous and incorporate into LL equation here we use fourier series to represent the periodic material variation into continous form. In the above equation, $g(x)$ represents continous form for the variation of saturation magnetization value at each points in the periodic ferromagnetic system having equal widths. By using different ferromagnetic materials to form a periodic ferromagnetic structure, form of the $g(x)$ changes accordingly.\\
From $\psi$ we obtain the components of magnetization,
\begin{equation}
  m_{x}(x,t) =  \left( \frac{\psi + \psi* }{2}\right)
\end{equation}
\begin{equation} 
 m_{y}(x,t)  =  \left( \frac{\psi - \psi*}{2i}\right)
\end{equation}
\begin{equation}
 m_{z}(x,t) = \left(1 -\lvert\Psi\rvert ^2\right)^{\frac{1}{2}} 
\end{equation}
The results indicates that the amplitude of the soliton solution depends on the nonlinearity coefficient $g(x)$, which means that the soliton can be spatially modulated and which admits several interesting spatial phenomena. \\ 
{\bf{Case (i):}}\\When $M_{S,A}$=$M_{S,B}$, then $g(x)$=1 it represents homogeneous ferromagneic system which is governed by standard nonlinear Schrodinger equation and admits soliton solution which propagates in a homogeneous background shown in Fig. (2).
\begin{figure}[htp]
\includegraphics[width=0.45\columnwidth]{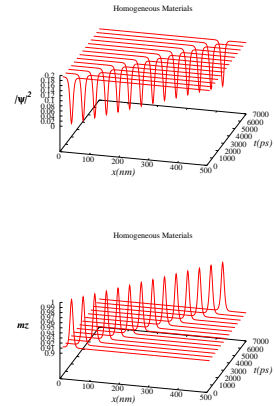}
\caption{Soliton propagates in a homogeneous background}
\end{figure}
\\
{\bf{Case (ii):}}\\When $M_{S,A}$$\neq$$M_{S,B}$, then\\
\begin{equation}
g(x)= 1 + \sum_{n=1} \left(\frac{M_{s,A}-M_{s,B}}{M_{S,av} n \pi}\right)(1-(-1)^{n}))sin(\frac{2n \pi x}{a}).
\end{equation}
\begin{equation}
\\
g(x)= 1 + \sum_{n=1} l_{n} sin(\frac{2n \pi x}{a}).
\end{equation}
Where $l_{n}$ is the control parameter which determines the nature of the magnonic crystal.\\
{\bf{Case (a):}}\\ 
\hspace{1cm} Consider a magnonic crystal system formed by periodic array of distinct ferromagnetic elements of iron,(Fe) and cobalt,(Co). The magnonic crystal is formed with lattice constant, a is 500nm by choosing the width of the each layer as 250nm. The spatial magnon density profile, $\lvert\Psi\rvert ^2$ and $m_{z}$ components in 3D and 2D of magnonic crystal for the combination of 250Fe/250Co are shown in Fig.(3a-3d). Its corresponding periodic applied magnetic field which is the condition for integrability from Eq.(41) is shown in Fig.(4). The excitation of magnetization in the form of spatially periodic localized modes is exists in the oscillatory background with structure similar to the form of spin Bloch waves[30].\\ 
%Lattice Constant of magnonic crystal a = 500nm~~~ 250Fe/250Co\\
%    \begin{tabular}{ | l |  p{5cm} | l | l | }
%    \hline
%    \bf{Material} & \bf{Saturation magnetization $M_{s}$ ($10^{6}$ A/m)} & \bf{Exchange length, $J_{ex}$ (nm)}\\ \hline
%    $M_{S,A}$, Fe & 1.752 & 3.30 \\ \hline
%    $M_{S,B}$, Co & 1.445 & 4.78\\ \hline
%\end{tabular}\\
%$M_{S,av}$= 1.598x$10^{6}$ A/m and $J_{ex,av}$= 4.04nm\\
Lattice Constant of magnonic crystal a=500nm 250Fe/250Co\\
\begin{small}
    \begin{tabular}{ | l |  p{2cm} | l | l | }
    \hline
    \bf{Material} & \bf{Saturation magnetization $M_{s}$ ($10^{6}$ A/m)} & \bf{Exchange length, $J_{ex}$ (nm)}\\ \hline
    $M_{S,A}$, Fe & 1.752 & 3.30 \\ \hline
    $M_{S,B}$, Co & 1.445 & 4.78\\ \hline
    \end{tabular}
\end{small}
$M_{S,av}$= 1.598x$10^{6}$ A/m and $J_{ex,av}$= 4.04nm\\
\begin{figure}[htp]
\begin{center}
\includegraphics[width=0.65\columnwidth]{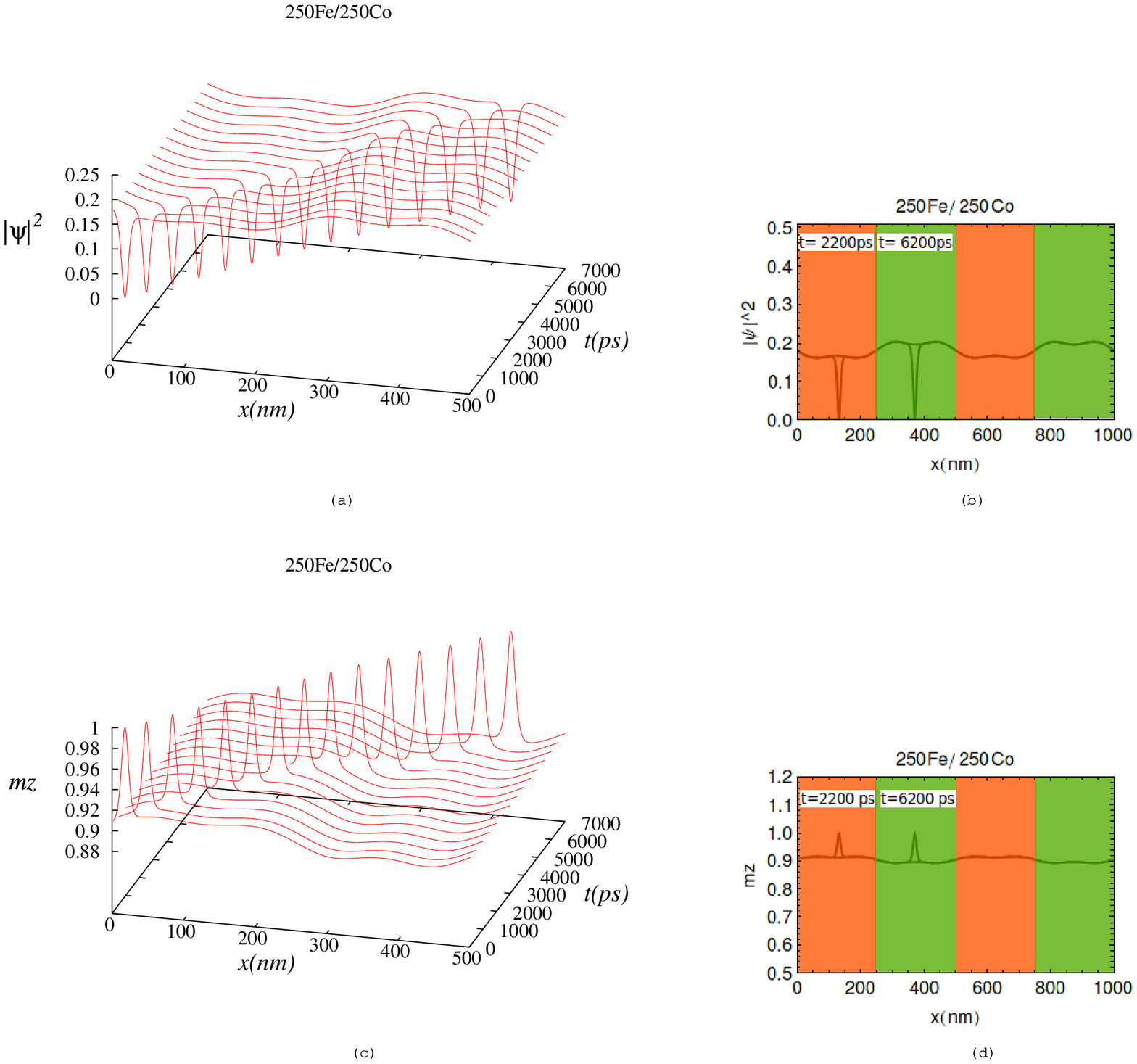}
\caption{(a) and (b) The spatial profile of magnon density $\lvert\Psi\rvert ^2$ in  the form of soliton on the background of spin Bloch waves in 3D and 2D. (c) and (d) $m_{z}$ component in 3D and 2D for 250Fe/250Co.}
\end{center}
\end{figure}
\begin{figure}[htp]
\includegraphics[width=0.35\columnwidth]{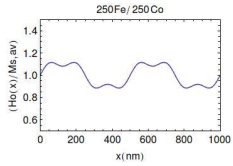}
\caption{Periodic applied magnetic field which is the condition for integrability from Eq. (41) for the combination of 250Fe/250Co. }
\end{figure}
{\bf{Case (b):}}\\
\hspace{1cm}Consider a magnonic crystal system  made of cobalt,(Co) and permalloy,(Py) with lattice constant, a=500nm by choosing each width as 250nm. The density profile, $\lvert\Psi\rvert ^2$ and $m_{z}$ components in 3D and 2D of magnonic crystal for the combination of 250Co/250Py shown in Fig.(5)  and its corresponding periodic applied magnetic field which is the condition for integrability from Eq.(41) are shown in Fig.(6). As in the previous case, the excitation of magnetization in the form of soliton has spatial period exists on the background of spin Bloch waves. The parameter $g(x)$ determines the nature of the magnetic crystal and periodic applied magnetic field which act as periodic potential for spin waves to satisfy the integrability condition. The stability nature of the soliton solution is depends on these parameters which are material dependent and its completely under our control. By choosing the ferromagnetic materials which forms the magnonic crystal of our interest, it is possible to tune the spatially modulated amplitude of the soliton. This spatially modulated amplitude soliton solutions with oscillatory background describes the nonlinear localized exicitation of magnetization in one dimensional magnonic crystal.\\ 
%Lattice Constant of magnonic crystal a = 500nm~~~ 250Co/250Py\\
 %   \begin{tabular}{ | l |  p{5cm} | l | l | }
  %  \hline
   % \bf{Material} & \bf{Saturation magnetization $M_{s}$ ($10^{6}$ A/m)} & \bf{Exchange length, $J_{ex}$ (nm)}\\ \hline
    %$M_{S,A}$, Co & 1.445 & 4.78 \\ \hline
    %$M_{S,B}$, Py & 0.860 & 7.64 \\ \hline
%\end{tabular}\\
%$M_{S,av}$= 1.152x$10^{6}$ A/m and $J_{ex,av}$= 6.21nm
Lattice Constant of magnonic crystal a=500nm 250Co/250Py\\
\begin{small}
    \begin{tabular}{ | l |  p{2cm} | l | l | }
    \hline
    \bf{Material} & \bf{Saturation magnetization $M_{s}$ ($10^{6}$ A/m)} & \bf{Exchange length, $J_{ex}$ (nm)}\\ \hline
    $M_{S,A}$, Co & 1.445 & 4.78 \\ \hline
    $M_{S,B}$, Py & 0.860 & 7.64 \\ \hline
    \end{tabular}
\end{small}\\
$M_{S,av}$= 1.152x$10^{6}$ A/m and $J_{ex,av}$= 6.21nm
\begin{figure}[htp]
\begin{center}
\includegraphics[width=0.65\columnwidth]{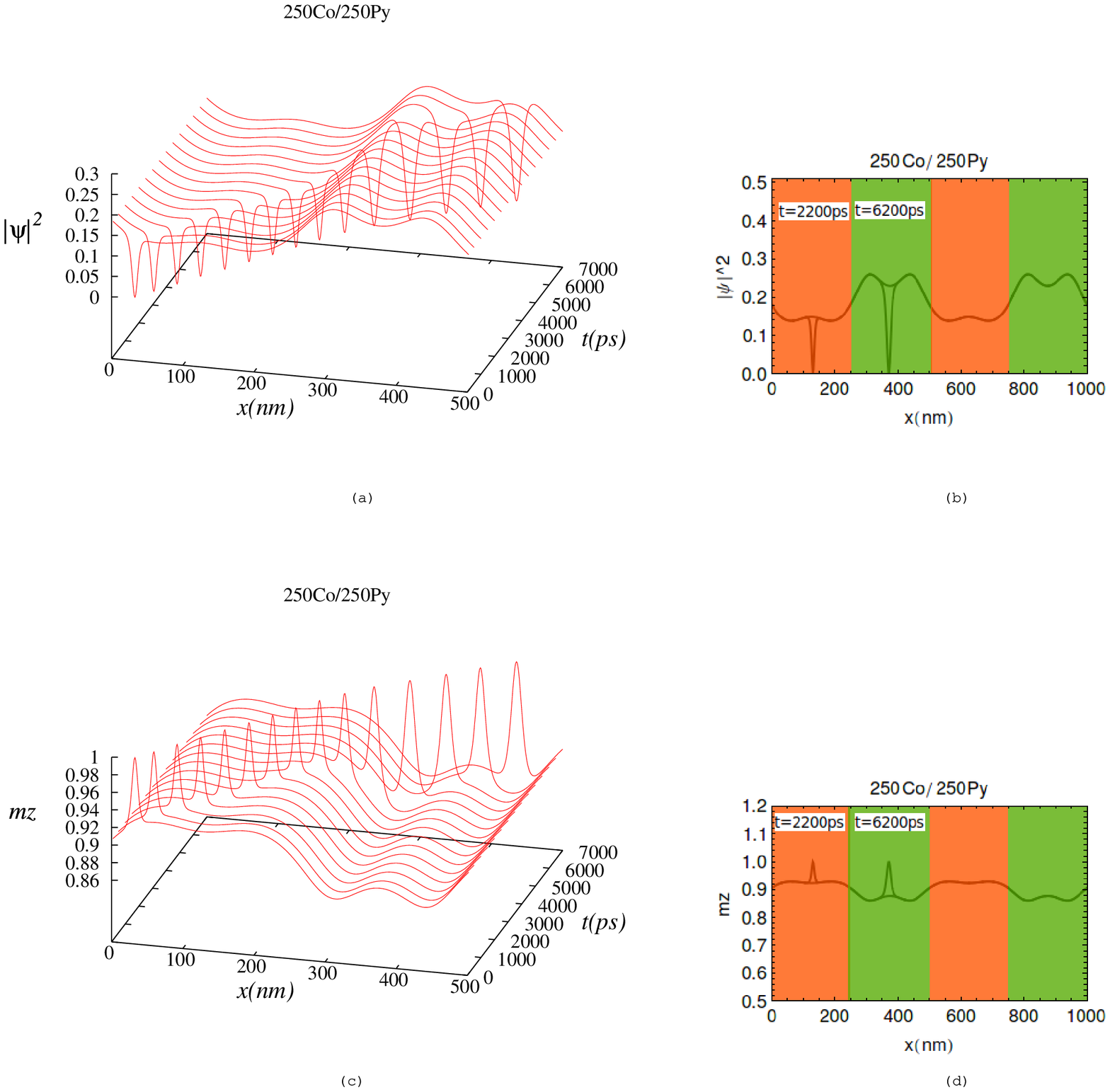}
\caption{(a) and (b) The spatial profile of magnon density $\lvert\Psi\rvert ^2$ in  the form of soliton on the background of spin Bloch waves in 3D and 2D. (c) and (d) $m_{z}$ component in 3D and 2D for 250Co/250Py.}
\end{center}
\end{figure}
\begin{figure}[htp]
\includegraphics[width=0.35\columnwidth]{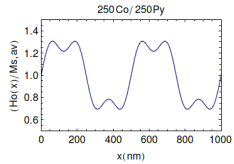}
\caption{Periodic applied magnetic field which is the condition for integrability from Eq. (41) for the combination of 250Co/250Py.}
\end{figure}
\newpage
\section*{V. Conclusion}
In this paper, by transforming the governing Landau-Lifshitz equation of one dimensional magnonic crystal into Variable coefficient Nonlinear Schrodinger equation(VCNLS), we have investigated the dynamics of magnetization in one dimensional magnonic crystal and with aid of painleve analysis we constructed the soliton solution exists on the oscillatory background with structure similar to the form of spin Bloch waves. Such solutions exists in certain constraint conditions on the coefficients of the VCNLS equation. The results shows that the amplitude of the soliton solution has spatial period on the background of spin Bloch waves. The spatial distribution of the soliton profile determined by the free parameter $g(x)$, which is depend on the saturation magnetization, $M_{s}$ values of the ferromagnetic materials which are used to form the magnonic crystal. From the case studies it is observed that by varying the parameter $g(x)$ nonlinear coefficient of the VCNLS equation for diiferent combination of the magnonic crystal, the form of the periodic applied magnetic field can be changed to satisfy the integrability condition and  which is act as the potential barrier for the spin waves and accordingly the desirable amplitude modulation of the soliton can be achieved.\\
\\
{\bf{References:}}\\
1. M. Lakshmanan, \emph{Phys. Lett.} {\bf{61A}}, 53 (1977).\\
2. M. Daniel, M. D. Kruskal, M. Laksmanan, and K. Nakamura, \emph{J. Math. Phys.} {\bf{33}}, 771(1992).\\
3. M. Daniel, K. Porsezian, and M. Lakshmanan, \emph{J. Math. Phys.} {\bf{35}}, 6498(1994).\\
4. M. Daniel and R. Amuda, \emph{Phys. Rev. B} {\bf{53}}, R2930 (1996).\\
5. M. Daniel, L. Kavitha, and R. Amuda, \emph{Phys. Rev. B} {\bf{59}}, 13774 (1999).\\
6. M. Daniel and L. Kavitha, \emph{Phys. Rev. B} {\bf{66}}, 184433 (2002).\\
7. P.B. He and W. M. Liu, \emph{Phys. Rev. B} {\bf{72}}, 064410 (2005).\\
8. M. Daniel and J. Beula, \emph{Phys. Rev. B} {\bf{77}}, 144416(2008).\\
9. Y. V. Kartashov, B. A. Malomed, and L. Torner, \emph{Rev. Mod. Phys.} {\bf{83}}, 247 (2011).\\
10. Shin H J, Radha R, Ramesh Kumar V  \emph{Phys.Lett.A} {\bf{375}} 2519(2011).\\
11. J. Y. Louis, E. A. Ostrovskaya, C. M. Savage, and Yu. S. Kivshar, \emph{Phys. Rev. A} {\bf{67}}, 013602 (2003).\\
12. T. Mayteevarunyoo, B. A. Malomed \emph{JOSA B}, {\bf{25}}, 1854 (2008).\\
13. A.B. Ustinov, N.Yu. Grigor'eva, B.A. Kalinikos, \emph{JETP Lett.} {\bf{88}}, 31 (2008).\\
14. A.B. Ustinov, B.A. Kalinikos, V.E. Demidov, S.O. Demokritov, \emph{Phys. Rev. B} {\bf{81}}, 180406 (2010).\\
15. A.V. Drozdovskii, M.A. Cherkasskii, A.B. Ustinov,N.G. Kovshikov, B.A. Kalinikos, \emph{JETP Lett.} {\bf{91}}, 16(2010).\\
16. M.A. Morozova, S.A. Nikitov, Yu.P. Sharaveski and S.E. Sheshukova, \emph{Acta Physica Polonica}, {\bf{121}}, 5 (2012).\\
17. P.B. He, G.N. Gu and A.L. Pan, \emph{Eur.Phys.J.B} {\bf{85}}, 119(2012).\\
18. M.Krawczyk and H. Puszkarski, Plane wave theory of three-dimensional magnonic crystals, \emph{Phys. Rev.B} {\bf{77}}, 054437 (2008).\\
19. M. Krawczyk, M. L. Sokolovskyy, J. W. Klos, and S. Mamica,  \emph{Advances in Condensed Matter Physics}, {\bf{2012}}, 764783 (2012).\\
20. A.M. Kosevich, B.A. Ivanov, A.S. Kovalev, \emph{Physics Reports}, {\bf{194}}, 117–238 (1990)\\
21. M. Lakshmananan and S. Rajasekar \emph{Nonlinear Dynamics: Integrability, Chaos and Patterns} Springer, Newyork, 2003.\\
22. A.V. Chumak, A.A. Serga, B. Hillebrands, and M.P. Kostylev, \emph{Appl. Phys. Lett.} {\bf{93}}, 022508 (2008).\\
23. M. E. Dokukin, K. Togo, and M. Inoue,\emph{J. Magn. Soc. Jpn.} {\bf{32}}, 103 (2008).\\
24. C. Bayer, M.P. Kostylew, and B. Hillebrands, \emph{Appl. Phys. Lett.} {\bf{88}}, 112504 (2006)
25. K. Wang, V. L. Zhang, H. S. Lim, S. C. Ng, M. H. Kuok, S. Jain, and A. O. Adeyeye, \emph{Appl. Phys. Lett.} {\bf{94}}, 083112 (2009).\\
26. K. Wang, V. L. Zhang, H. S. Lim, S. C. Ng, M. H. Kuok, S. Jain, and A. O. Adeyeye, \emph{ACS Nano} {\bf{4}}, 643 (2010).\\
27. C.Ozemir and F. Gungor \emph{Rev. Math. Phys.} {\bf{24}}, 1250015 (2012).\\
28. U. Al Khawaja, \emph{Physics Letters A} {\bf{373}}, 2710-2716(2009).\\
29. S. Rajendran, P. Muruganadam, M. Laxmannan, \emph{Physica D}, {\bf{239}}, 366-386(2010).\\
30. C.S. Lin, H.S. Lim, Z.K. Wang, S.C. Ng, and M.H. Kuok, \emph{Appl. Phys. Lett.}, {\bf{98}} 022504 (2011).
\end{document}